\documentclass[11pt, a4paper]{article}
\usepackage{graphicx}
\usepackage{jheppub}
\usepackage{epsfig}
\usepackage{psfrag,verbatim}

\def\caln         {{\cal N}}
\def\calo         {{\cal O}}

\def\zet          {{\mathbb Z}}

\def\sqr#1#2{{\vcenter{\vbox{\hrule height.#2pt
 \hbox{\vrule width.#2pt height#1pt \kern#1pt
 \vrule width.#2pt}\hrule height.#2pt}}}}

\def\a{\alpha}

\def\dim{{\rm dim}}

\newcommand{\beq}{\begin{equation}}
\newcommand{\eeq}{\end{equation}}
\newcommand{\beqa}{\begin{eqnarray}}
\newcommand{\eeqa}{\end{eqnarray}}
\newcommand{\beqar}{\begin{eqnarray*}}
\newcommand{\eeqar}{\end{eqnarray*}}

\renewcommand{\eqref}[1]{(\ref{#1})}

\newcommand{\ie}{{\it i.e.,}\ }

\newcommand{\abs}[1]{\lvert#1\rvert}
\providecommand{\abs}[1]{\lvert#1\rvert}

\def\a{\alpha}

\def\c{\chi}
\def\k{\kappa}

\catcode`\@=12

\title{\bf Eling-Oz Formula for Exotic Hairy Black Holes }

\author{
Alexander Patrushev$ ^{1}$\\[0.4cm]
\it $ ^1$Department of Applied Mathematics\\
\it University of Western Ontario\\
\it London, Ontario N6A 5B7, Canada\\
}

\emailAdd{apatrush@uwo.ca}
\arxivnumber{hep-th: 1107.2385}

\abstract{
We checked Eling-Oz formula \cite{eo} for the bulk viscosity of the holographic fluid dual to the exotic black holes \cite{bp1}. Initially, Eling and Oz argued that the formula is valid
in the high temperature and adiabatic limits. In \cite{boz} the validity of the formula for $\caln=2^*$ plasma and cascading gauge theory was pushed forward for arbitrary temperatures. 
We successfully verified the formula with the computations of the bulk viscosity \cite{crit,bp2010} for a wide range of the temperatures. Moreover, it correctly reproduces the critical behavior 
in the vicinity of the critical point,
where the bulk-to-shear viscosity diverges.}

\begin{document}

\maketitle

\section{Introduction and Summary}

A new formula for bulk-to-shear viscosity of strongly coupled gauge theory plasma was proposed by Eling and Oz (EO) \cite{eo}. The wide class of the gauge theories are dual (in the context of AdS/CFT \cite{juan})
to the following  $(d+1)$ gravitational action
\begin{equation}
S= \frac{1}{16\pi} \int \sqrt{-g} d^{d+1} x \left(R - \frac{1}{2} \sum_i (\partial \phi_i)^2 - V(\phi_i)\right) + S_{gauge}.
\label{action}
\end{equation}
They used the null focusing (Raychaudhuri) equation describing the evolution of the horizon entropy, which is equivalent to the viscous fluid entropy balance law.
In the absence of chemical potentials for the conserved charges, the formula for the bulk viscosity of the plasma dual to \eqref{action} takes the following form
\begin{equation}
\frac{\zeta}{\eta} = \sum_i c_s^4 T^2 \left( \frac{d \phi^{H}_i}{dT}\right)^2 \ ,
\label{eo1}
\end{equation}
where $\phi_i^H$ are the  scalar field values evaluated at the horizon of the black brain, $T$ is the temperature of plasma dual to the black brane,
$c_s$ is the speed of sound waves in plasma.
In the same paper the EO formula was verified for a large number of gauge theories dual to string theory at high temperature limit \cite{v1,v2,v3,v4} and 
some phenomenological models of gauge/gravity correspondence \cite{vp1,vp2}.

The expression for the bulk-to-shear viscosity employs the values of scalar fields only at the horizon. It is an intriguing result as the bulk viscosity
in general depends on the energy scale; the boundary data is essential to capture microscopic scales of the theory. That is in contrast with the universality of the shear
viscosity calculations \cite{u1,u2,u3}.
In \cite{boz} the validity of \eqref{eo1} was extended for cascading gauge plasma \cite{c1,c2} and $\caln=2^*$ gauge theory plasma \cite{n21,bp2,cr1} for all the temperatures.

The correctness of the EO formula for the phenomenological models of gauge/gravity correspondence was also verified in \cite{agpr}. Particularly, bulk viscosity obtained from the
Gubser, Pufu and Rocha (GPR) formula for the GPR model \cite{gpr} and the Improved Holographic QCD model \cite{imqcd} coincides with the EO formula.
The essential feature of the GPR formula (extracted from the holographic Kubo formula) is that it is suitable only for the models with one gravitational scalar field acting as the new radial coordinate. 
The exotic black hole model is the example of the phenomenological model with several gravitational scalar fields.

We checked the validity of the Eling-Oz formula analytically for the exotic black holes in the high-temperature (conformal) limit. The formula is correct for the intermediate temperatures,
the vicinity of the phase transition and  
for the temperatures up to $\frac{m}{T}<2.75$, where $m$ is the mass associated with breaking of the conformal symmetry. \footnote{Original calculations of the bulk viscosity were done in \cite{crit,bp2010}.}
The correctness of the formula for exotic black holes extends 
the number of models for which the EO formula is valid for all energy scales.  It would be interesting to explain this kind of universality of \eqref{eo1} (see also footnote [4] in the text).  
While preparing this manuscript, I learned that the authors of \cite{eo} proved such a universality of the transport coefficients of the holographic plasmas \cite{eoscreen}. 
They showed that the transport coefficients depend on the boundary conditions, but they are independent of the RG running from UV to IR.

\section{Bulk viscosity for the exotic hairy black holes}

The exotic black hole model is defined by the following effective (3+1)-dimensional gravitational action \cite{bp1,phd}:
\begin{equation}
S_4=\frac{1}{2\kappa^2}\int dx^4\sqrt{-\gamma}\left[R+6 
-\frac 12 \left(\nabla\phi\right)^2+\phi^2-\frac 12 \left(\nabla\chi\right)^2-2\chi^2-g \phi^2 \chi^2
\right]\,,
\label{s4}
\end{equation}
where $g$ is a coupling constant.\footnote{In numerical analysis we set $g=-100$.}
Note that $\phi$ induces a relevant deformation of the dual CFT by an operator $\calo_r$ and
$\chi$ is associated with an irrelevant operator $\calo_i$ in the dual gauge theory. The last term in \eqref{s4} involves mixing of $\calo_i$ with $\calo_r$ under RG dynamics.
The central charge of the UV fixed point is defined as \cite{bp1}
\begin{equation}
c=\frac{192}{\k^2}\,,
\label{c}
\end{equation} 
This central charge should be understood as a measure of the degrees of freedom in the CFT,
which is defined thermodynamically or via two-point correlation functions \cite{kov}.
We demand the solution to be  $AdS_4$ asymptotically  with translationary invariant horizon. For this purpose only the normalizable mode of $\calo_i$ is nonzero near the boundary.

The background geometry is defined as 
\begin{equation}
ds_4^2=-c_1(r)^2\ dt^2+c_2(r)^2\ \left[dx_1^2+dx_2^2\right]+c_3(r)^2\ dr^2\,,\qquad \phi=\phi(r)\,,\qquad 
\chi=\chi(r)\,,
\label{background}
\end{equation}
where $r\rightarrow\infty$ corresponds to the AdS boundary.
Then one can introduce a new radial coordinate $x$ as follows
\begin{equation}
1-x\equiv \frac{c_1(r)}{c_2(r)}\,,
\label{xdef}
\end{equation} 
so that $x\to 0$ corresponds to the AdS boundary, and $y\equiv 1-x\to 0$ 
corresponds to a horizon asymptotic.
Afterwards, we introduce $a(x)$ as 
\begin{equation}
c_2(x)=\frac{a(x)}{(2x-x^2)^{1/3}}\,,
\label{defa}
\end{equation}
The equations of motion (EOMs) obtained from \ref{s4}, with the background 
ansatz \eqref{background}, define the following expansion of the model parameters 
\begin{equation}
\begin{split}
a=&\a\left(1-\frac{1}{40}\ p_1^2\ x^{2/3}-\frac{1}{18}\ p_1 p_2\ x+\calo(x^{4/3})\right)\,,\\
\phi=&p_1\ x^{1/3}+p_2\ x^{2/3}+\frac{3}{20} p_1^3 x+\calo(x^{4/3})\,,\\
\c=&\c_4\left(x^{4/3}+\left(\frac 17 g -\frac{3}{70}\right)p_1^2\ x^2+\calo(x^{7/3})\right)\,,
\end{split}
\label{boundary}
\end{equation}
near the boundary $x\to 0_+$, and 
\begin{equation}
a=\a\left(a_0^{h}+a_1^h\ y^2+\calo(y^4)\right)\,,\qquad \phi=p_0^h+\calo(y^2)\,,\qquad  \c=c_0^h+\calo(y^2)\,,
\label{horizon}
\end{equation}
near the horizon. 
Up to the overall scaling factor $\a$  the thermodynamics of the black branes can be uniquely specified
with 3 UV coefficients $\{p_1,p_2,\c_4\}$ and 4 IR coefficients 
$\{a_0^h,a_1^h,p_0^h,c_0^h\}$. 
 
We use the integral of motion \cite{phd}
\begin{equation}
 (a_0^h)^2\sqrt{\frac{(6a_0^h)^3(6+(p_0^h)^2-2(c_0^h)^2-g(p_0^h)^2 (c_0^h)^2)}{3a_1^h+a_0^h}}=6,
\end{equation}
which arises after integration of the EOMs, 
to find the temperature $T$ and the entropy density $s$ of the black brane solution 
\eqref{background}:
\begin{equation}
T=\frac{3\alpha}{4\pi (a_0^h)^2}\,,
\label{tbh}
\end{equation}
\begin{equation}
\hat{s}\equiv \frac{384}{c}\ s= 4\pi\a^2\ (a_0^h)^2\,,
\label{sbh}
\end{equation}
In the dual picture $p_1$ can be interpreted as the coupling of the operator $\calo_r$, $p_2$ as the expectation value of $\calo_r$ and $\chi_4$ as $<\calo_i>$ (see \cite{KW} for the argumentation).
Without the loss of generality, we choose the model with $\dim[\calo_r]=2$. Which means the combination $p_1 \a$  should be fixed.

For a given set of $\{\a,p_1\}$ there is a discrete set of the remaining parameters 
\[
\{a_0^h,\ p_0^h, c_0^h\}
\]
allowing us to characterize the thermodynamics of black branes suitable for Eling-Oz formula. One solution with  $c_0^h=0$
describes the black brane without the condensate of the $\c$ field. All the other solutions 
have  $c_0^h\ne 0$ and describe the ``exotic black branes'' \cite{bp1}. This model is interesting because the transition occurs at the high temperatures (rather than the low temperatures).
That is the irrelevant operator $\calo_i$ obtains nonzero vacuum expectation value for $T>T_c$ , spontaneously breaking a discrete $\zet_2$ symmetry of the model.
Also, it was shown in \cite{bp2010} that all the exotic black branes contain a tachyonic quasinormal mode. Thus, they are dynamically unstable but thermodynamically stable, thereby, violating 
the correlated stability conjecture \cite{gm1,gm2,csc}.

For the exotic black hole model formula \eqref{eo1} takes the following simple form
\begin{equation}
\frac{\zeta}{\eta}\bigg|_{EO}=c_s^4\tau^2\left(\left(\frac{dp_0^h}{d\tau}\right)^2+\left(\frac{dc_0^h}{d\tau}\right)^2\right)\,,
\label{eo}
\end{equation}
where $c_s$ is a speed of sound defined by
\begin{equation}
 c_s^2=\frac{d(\ln T)}{d(\ln s)},
\end{equation}
and $\tau$ is an inversed dimensionless temperature, e.g.\ $\tau=\frac{T_c}{T}$ or $\tau=\frac{\alpha p_1}{T}$.
Further, we will check the validity of the formula for the different temperature regimes.

\subsection{Explicit analytical check of \eqref{eo1} in the conformal limit of a symmetric phase}

\begin{figure}[t]
\begin{center}
\psfrag{chi0}{{$p_0^h$}}
\psfrag{TcT}{{$\frac{T_c}{T}$}}
\psfrag{xi}{{$\frac{\zeta}{\eta}$}}
\includegraphics[width=2.9in]{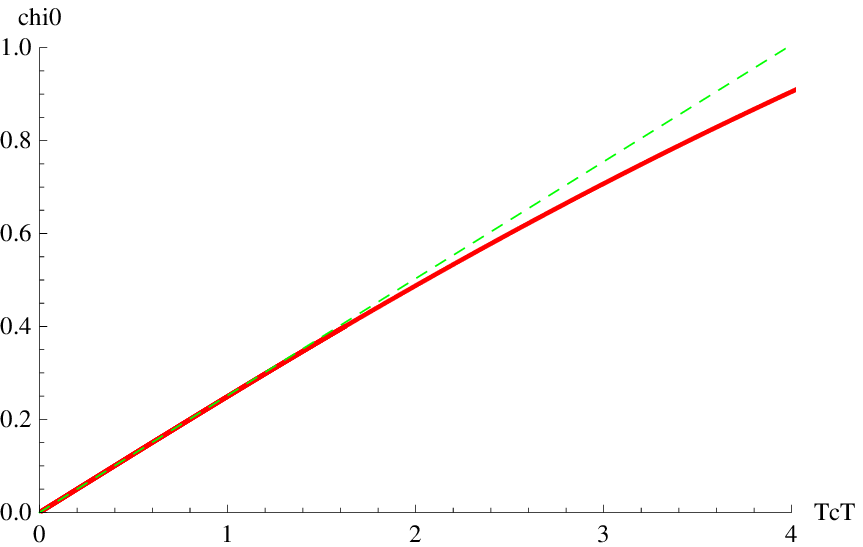}
\includegraphics[width=2.9in]{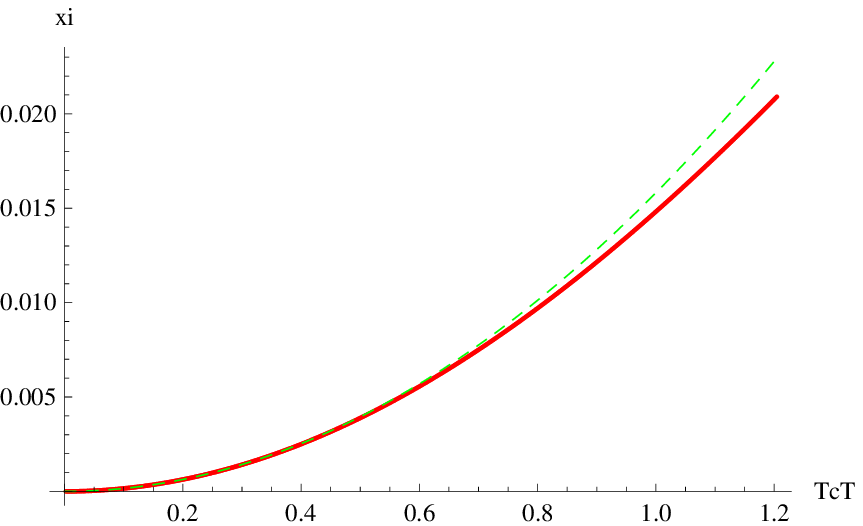}
\end{center}
  \caption{(Colour online)
Comparison of a scalar field at the horizon and the bulk viscosity computed using quasinormal modes with the high-temperature limit \cite{bp1}.
The dashed green lines represent the conformal limit.
 } \label{figure1}
\end{figure}

In this section we briefly repeat the main results for the thermodynamics of the model in the high temperature limit as presented in \cite{phd}. 
These results can be readily applied to supply evidence of \eqref{eo} in the conformal limit. We demonstrate this below.
For the sake of simplicity, we consider the symmetric phase only ($\chi=0$). If we introduce the  small 
deformation parameter $\delta$ such that
\begin{equation}
 \delta=\frac{m}{T}<<1,
\end{equation}
where $m$ is the mass associated with the deformation of CFT. 
Then one can expand the scalar field $\phi$ in the EOMs to leading order in $\delta$ as 
\footnote{It is a natural small parameter to expand the solutions of EOMs defining the values of the scalar fields.}
\begin{equation}
 \phi(y)=\delta\tilde\phi(y).
\end{equation}
Afterwards, we solve the EOMs demanding regularity of the scalar field at the horizon and the first law of thermodynamics to fix the integration constants.
\footnote{In general, $p_2$ --- the expectation value of $\calo_r$ is connected to the horizon data through the boundary conditions, regularity of the solution of the EOMs at the horizon and boundary. 
Particularly, in the conformal limit $p_2\propto p_0^h.$
This can be the qualitative argument that the EO formula captures the UV data from the boundary.}

The expansion of the scalar field $\phi$ near horizon \eqref{horizon} assumes that
\begin{equation}
 p_0^h=\delta.
\label{ph}
\end{equation}
Eventually, it leads to the following expression for the speed of sound in the conformal limit
\begin{equation}
 c_s^2=\frac{1}{2}-\frac{\sqrt 3}{8 \pi} \delta^2+\calo(\delta^4).
\label{cs}
\end{equation}
The results for the bulk viscosity in the conformal limit were discussed in \cite{crit}.
For a  symmetric phase at the high temperatures we have 
\begin{equation}
\frac{\zeta}{\eta}\bigg|_{ordered}=\frac{2\pi}{\sqrt{3}}\left(\frac 12 -c_s^2\right)+\calo\left(\left(\frac12-c_s^2\right)^2\right)\,.
\label{hight}
\end{equation}
A a result, the bulk viscosity should be proportional to the square of the deformation parameter, \ie
\begin{equation}
 \frac{\zeta}{\eta}=\frac{1}{4}\delta^2.
\label{confbulk}
\end{equation}
If we substitute \eqref{ph} and \eqref{cs} into \eqref{eo}  we recover exactly the same relation ($\tau=\delta$ in this case).
This justifies the validity of \eqref{eo} in the conformal limit.

One can see that in the conformal limit the value of the scalar field at the horizon is proportional to $\frac{m}{T}$. 
In Fig. \ref{figure1} we use a linear approximation for the scalar field at the horizon to establish the connection between $\delta$ and $\frac{T_c}{T}$.
In addition, on the same figure we compare the bulk viscosity evaluated from the sound waves attenuation coefficient with 
analytical result \eqref{confbulk} in the conformal limit.

\subsection{Comparison of \eqref{eo1} away from criticality for symmetric and unstable phases}
Now it is possible to check the formula for an intermediate temperature regime. The original calculations of the thermodynamics and the bulk viscosity were done in \cite{bp1,crit}. 
We use a cubic spline approximation for the values of the scalar fields at the horizon to get their derivatives with respect to the inversed temperature.
Then we compare the validity of the Eling-Oz formula with respect to two arguments $\delta$ and $\frac{T_c}{T}$.

\begin{figure}[t]
\begin{center}
\psfrag{abserr}{{$\abs{\frac{{\zeta}/{\eta}|_{EO}}{{\zeta}/{\eta}}-1}$}}
\psfrag{TcT}{{$\frac{T_c}{T}$}}
\psfrag{del}{{$\delta$}}
\includegraphics[width=2.9in]{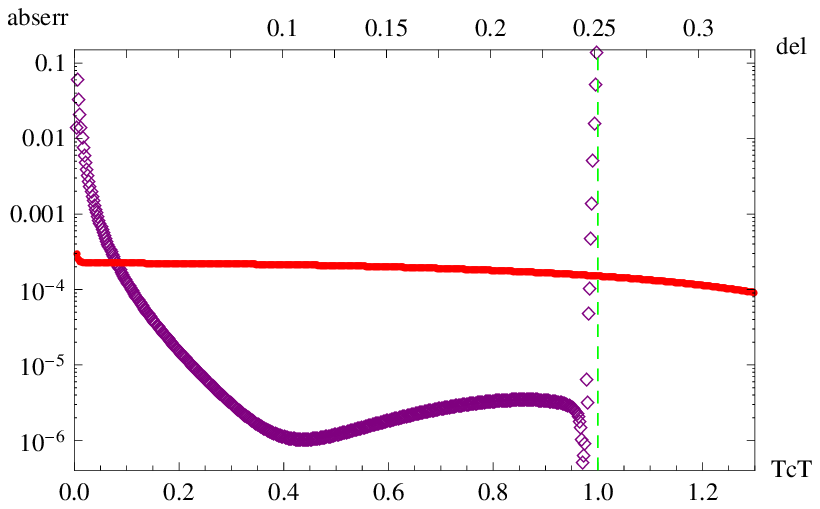}
\includegraphics[width=2.9in]{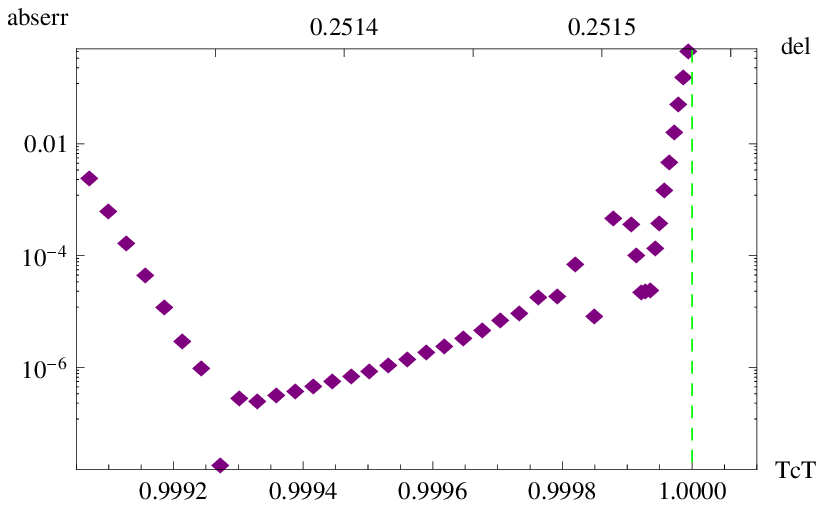}
\end{center}
  \caption{(Colour online)
Comparison of the EO formula for exotic black holes with the computations using quasinormal modes \cite{crit} far from criticality (the left plot) and in the vicinity of the critical point (the right plot).
Red circles are for the  ``ordered'' (symmetric) phase, purple diamonds are for the ``disordered'' (broken) phase.
The dashed vertical green line represents the critical point of the theory $T=T_c$.
 } \label{figure2}
\end{figure}

Fig. \ref{figure2} illustrates the absolute value of relative error between the bulk viscosity obtained from quasinormal mode method and the Eling-Oz formula. The error is slightly big
for large temperatures due to small values of the scalar fields (which increases numerical errors). But we have the analytic results from the previous section for this region of temperatures. 
The agreement is also worse in the vicinity of the critical point (due to large values of $\frac{dp_0^h}{d\tau}$). In the next subsection we will improve the agreement in the 
critical regime.

\subsection{Comparison of \eqref{eo1} at criticality}

The critical behavior of the exotic black branes was discussed in \cite{crit}.
The peculiar thing in the model is that the bulk-to-shear viscosity diverges in the broken phase at criticality.
One can use more detailed data to check the formula close to criticality as it is done in the Fig. \ref{figure2}. Alternatively, we can
proceed the semi-numerical analysis.

\begin{figure}[t]
\begin{center}
\psfrag{chi0}{{$(c_0^h)^2$}}
\psfrag{TcT}{{$\frac{T_c}{T}$}}
\psfrag{lnxi}{{$\ln \left(\frac{\zeta}{\eta}\right)$}}
\psfrag{lnchi}{{$\ln \left(c_0^h\right)$}}
\includegraphics[width=2.9in]{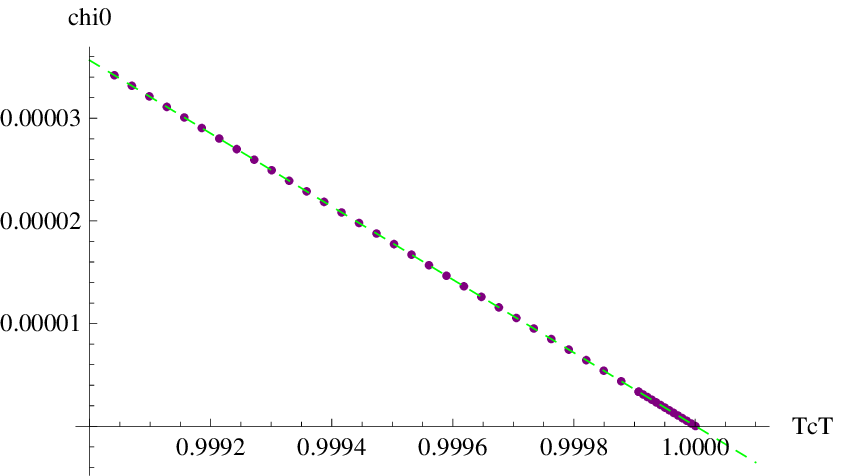}
\includegraphics[width=2.9in]{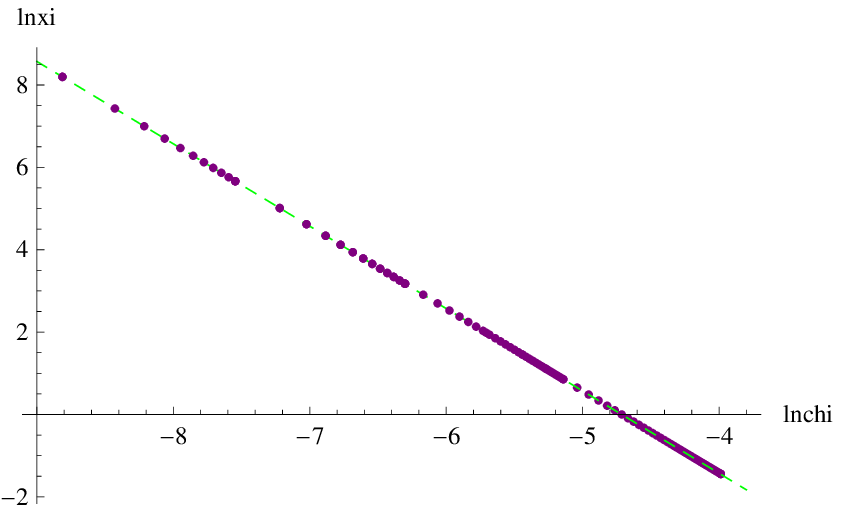}
\end{center}
  \caption{(Colour online)
The square of a scalar field at the horizon $c_0^h$ as function of the reduced temperature $\frac{T_c}{T}$.
The dashed green line is a linear fit to $(c_0^h)^2$
The right plot shows the bulk viscosity dependence on $c_0^h$.
The dashed green line represents a linear fit to the log-log data.
 } \label{figure3}
\end{figure}

In \cite{crit} the authors constructed an exotic model of the second order transition in $d=3$ at finite temperature and zero chemical potentials. The corresponding conformal field theory in $2+1$
dimensions is deformed by a relevant operator $\calo_r$. The expectation value  of $\calo_i$ acts as the order parameter of the phase transition and it scales as $\abs{t}^{1/2}$ 
in the vicinity of the critical point, where $t=\frac{T-T_c}{T_c}$.
In Fig. \ref{figure3} we plot $\ln\left(\frac{\zeta}{\eta}\right)$ versus $\ln \left(c_0^h\right)$ with a dashed green line which fits the data
\begin{equation}
 y=-1.999(8) x-9.427(8).
\end{equation}
It is clear from the Fig. \ref{figure3} that parameter $c_0$ has the same critical exponent as $<\calo_i>$
\begin{equation}
  c_0\propto \abs{t}^{1/2}.
\end{equation}
For the speed of sound and bulk viscosity it was shown that 
\begin{equation}
 c_s\propto \abs{t}^0\,,\qquad \frac{\zeta}{\eta}\bigg|_{disordered}\propto \abs{t}^{-1}.
\end{equation}
Whereas, formula \eqref{eo1} suggests that at criticality
\begin{equation}
 \frac{\zeta}{\eta}\bigg|_{EO}=c_s^4 T_c^2\left(\frac{c_0}{T-T_c}\right)^2\propto \abs{t}^{-1},
\end{equation}
which confirms the correctness of the EO formula at criticality.
Also, we expect that the formula will be valid for other symmetry-broken phases.

\subsection{Validity of \eqref{eo1} for the low temperatures}

\begin{figure}[t]
\begin{center}
\psfrag{abserr}{{$\left(\frac{{\zeta}/{\eta}|_{EO}}{{\zeta}/{\eta}}-1\right)$}}
\psfrag{TcT}{{$\frac{T_c}{T}$}}
\psfrag{del}{{$\delta$}}
\includegraphics[width=4in]{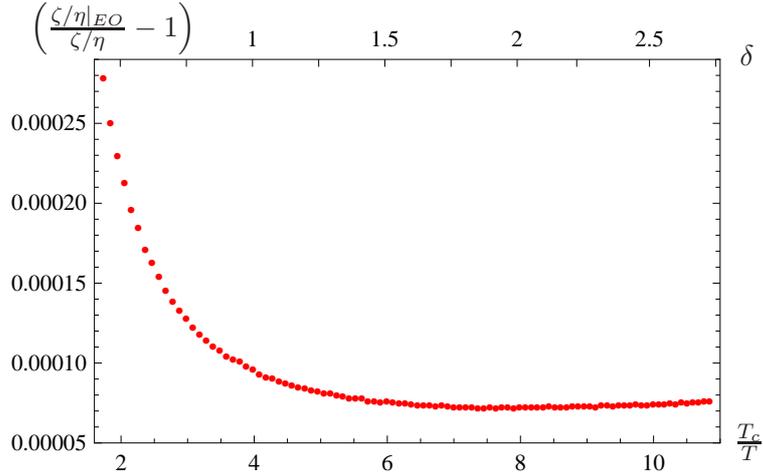}
\end{center}
  \caption{
Comparison of the EO formula for symmetric phase with the computations using quasinormal modes for the low temperatures.
 } \label{figure4}
\end{figure}

Let us proceed to the check of the Eling-Oz formula in the case when the stable phase is driven into the low temperature regime. 
First note that data for the thermodynamic parameters is more discrete, while the number of points for viscosity is significantly reduced (100 points altogether). 
Furthermore, Fig. \ref{figure4} presents a plot of the relative error with respect to $\delta$ and $\frac{T_c}{T}$. One can readily see that formula (2.11) is valid up to $\delta<2.75$. 
Therefore, combining the results of \cite{boz}, we expect that the formula is valid for the whole range of the temperatures.

\section*{Acknowledgments}
I would like to thank Alex Buchel for the formulation of the problem and helpful discussions. 
I am grateful to Chris Pagnutti for our many fruitful discussions and for providing numerical data and a patient explanation of the numerical techniques used in \cite{bp1} and \cite{crit}. 
Many thanks go to Michael Smolkin for reading the manuscript and improving the explanation. Also, Ildar Rakhmatulin made a helpful suggestion.

\end{document}